\documentclass[10pt]{article}
\usepackage{aas2pp4}
\usepackage{aassize}

\newcommand{\et}{{\it et~al.}}

\begin{document}
\title{X-ray Emission from the Guitar Nebula}
\vspace{-0.4cm}

\author{Roger W. Romani\altaffilmark{1}}
\affil{Dept. of Physics, Stanford University, Stanford CA  94305-4060}
\authoraddr{Dept. of Physics, Stanford University, Stanford CA  94305-4060}
\vspace{-0.6cm}

\author{James M. Cordes\altaffilmark{2}}
\affil{Astronomy Dept. and NAIC, Cornell University, Ithaca, NY 14853}
\authoraddr{Astronomy Dept., Cornell University, Ithaca, NY 14853}
\vspace{-0.6cm}

\author{and I.-A. Yadigaroglu}
\affil{Dept. of Physics, Stanford University, Stanford CA  94305-4060}
\authoraddr{Dept. of Physics, Stanford University, Stanford CA  94305-4060}
\vspace{-0.6cm}

\altaffiltext{1}{Alfred P. Sloan Fellow; rwr@astro.stanford.edu} 
\altaffiltext{2}{cordes@astrosun.tn.cornell.edu} 

\vspace{-.5cm}
\begin{abstract}
	We have detected weak soft X-ray emission from the Pulsar Wind
Nebula trailing the high velocity star PSR 2224+65 (the `Guitar Nebula').
This X-ray flux gives evidence of $\gamma \sim 10^7$ eV particles in the
pulsar wind and constrains the properties of the post-shock flow. The X-ray 
emission is most easily understood if the shocked pulsar wind is
partly confined in the nebula and if magnetic fields in this zone can grow 
to near equipartition values.
\end{abstract}

\vspace{-0.2cm}

\twocolumn

\section{Introduction}

	Identifying the spindown products of rotation-powered pulsars
is a long standing problem
in astrophysics. Recent observations of high energy emission from young pulsars
(eg. Fierro 1996) and improvement in emission models (eg. Romani 1996) show that
up to a few $\times 10$\% of the spindown power can be
ascribed to observed photons. In most cases, the remaining power in 
particles and unobserved radiation still needs to be identified. 
The wind of energetic particles and fields emitted from spin-powered pulsars 
is most easily
observed when it is confined. Such post-shock, or `plerionic'  emission has
long been associated with sources such as the Crab and Vela pulsars and with
similar filled-center remnants such as 3C58, presumably containing 
young pulsars.

	The discovery of bow shock nebulae around a number of pulsars has 
provided a further opportunity to study pulsar winds. These nebulae generally
occur when the pulsar is either long-lived, with a low magnetic field B and 
a short spin period, or when it has a large space velocity; these properties 
allow escape from the pulsar's parent supernova remnant while it retains
substantial spindown power. The pulsar wind then shocks against
the ambient ISM. These objects are most 
prominent as H$\alpha$ nebulae, first discovered around the millisecond 
pulsar PSR 1957+20 (Kulkarni and Hester 1988, see also Aldcroft, Romani 
and Cordes 1992). A similar bow shock is seen around the 5 ms pulsar
PSR J0437-4715 (Bell, Bailes and Bessel 1993) and Cordes, Romani and Lundgren 
(1993, CRL) have detected an H$\alpha$ nebula around the
high velocity, slow spin pulsar PSR 2224+65.

	High energy observations of these pulsar wind nebulae (PWNs) can 
place key constraints on the pulsar particle output. Models of the Crab nebula 
(Kennel \& Coroniti 1984), for example indicate that the bulk of the spindown 
\newpage
\mbox{}
\begin{minipage}[l]{3in}
\vspace{3.55in}
\end{minipage}

\hskip -1.25em 
power is carried off in an $e^\pm$ wind with $\gamma\sim 10^{6.5}$. Arons 
and Tavani (1993) described the expected high energy emission from wind shocks 
in the PSR 1957+20 system, focusing on a wind model including a massive 
baryonic component. Using ROSAT {\it Position Sensitive Proportional Counter}
(PSPC) observations of PSR 1957+20, Kulkarni 
et al. (1992) were able to extract general constraints on the $e^\pm$
content of the pulsar wind. A trail of soft X-ray
emission behind the nearby pulsar PSR1929+10, discovered in ROSAT PSPC
data by D. Helfand and described by Wang, Li and Begelman (1993), is also
a likely product of pulsar wind-ISM interaction. Finally the interaction
of pulsar winds with companion stellar winds can also produce a bow shock
geometry. For example, PSR 1259-63 in an eccentric orbit around a Be companion
provides the opportunity of studying the pulsar wind interactions under
changing conditions (Arons and Tavani 1994). ASCA observations of 
non-thermal X-rays (see Hirayama 1996, and references therein) have in
particular provided important constraints on the particle energy and 
luminosity.

	We describe here ROSAT {\it High Resolution Imager} (HRI) observations
of the PWN trailing PSR 2224+65, the `Guitar nebula'. This is followed
by a description of the bow shock flow geometry and a discussion of the origin
of the observed X-ray flux.  The Guitar nebula is particularly 
interesting because the pulsar itself is relatively normal, but the high 
velocity and peculiar shock geometry of the system provide opportunities 
to probe the pulsar wind interactions under conditions quite different 
than those found in other PWN. In particular, the small bow shock 
scale allows soft X-ray constraints on 
electron energies similar to those inferred for the Crab pulsar.

\section{ROSAT HRI flux}\label{xrays}

	We proposed ROSAT HRI observations of PSR 2224+65 to detect emission in
the shocked wind and ISM and to constrain the physics of the post-shock flow. 
Exposures were obtained in 7 observation intervals during 
July 13 -- 25, 1994.  With 36,894s of 
dead-time corrected exposure centered on PSR 2224+65 we were able to detect 
coronal emission from $\sim 10$ bright stars in the field. Cross 
correlation with optical frames registered to deep H$\alpha$ images showing 
the Guitar nebula allowed us to `boresight' the HRI data. Three 
{\it Hubble Space Telescope} (HST) guide stars
(GSC 472-144, -188, -232) were detected with standard processing in the center
of the HRI frame and were used for final alignment of the X-ray and optical
images to $< 1^{\prime\prime}$. This allowed a very sensitive search for 
emission from PSR 2224+65 and the Guitar Nebula.  No X-ray emission was 
detected 
from PSR 2224+65 itself, with the 2 counts detected within 4$^{\prime\prime}$ 
of the pulsar position entirely consistent with the 2.02 counts expected from
the mean background rate.  However an X-ray excess was apparent in the 
vicinity of the H$\alpha$ nebula. We have defined an aperture of half 
width $4^{\prime\prime}$ following the outline of the edge-brightened 
Guitar shape. If the X-ray emission is confined to the limb, then this
contour will contain 90\% of the source counts; if the bubble interior
has X-ray emission, the omitted flux will be slightly larger. This contour
contains 31 counts at the Guitar position
in excess of the 58 counts expected from the mean background 
rate (Figure 1). This is formally a $4.1\sigma$ detection.  To check the 
significance, we summed counts in our nebula-shaped aperture at various 
independent positions in the HRI image, after applying vignetting 
corrections and excluding regions where significant 
point sources (generally identified stars) were found in the standard PROS 
processing. The histogram of the resulting excess counts shows that the 
true Guitar position has the largest excess.  Estimating the envelope of the 
count fluctuations again gives a significance slightly in excess of $4\sigma$. 
The Guitar nebula thus emits soft X-rays with a count rate of
$9.6 \times 10^{-4}$counts/s in the ROSAT band. The flux seems to follow the
limb of the Guitar, concentrating in the forward facing parts of the nebula, but
starting significantly behind the pulsar itself. 

\begin{figure}[GuitarX]
\plotone{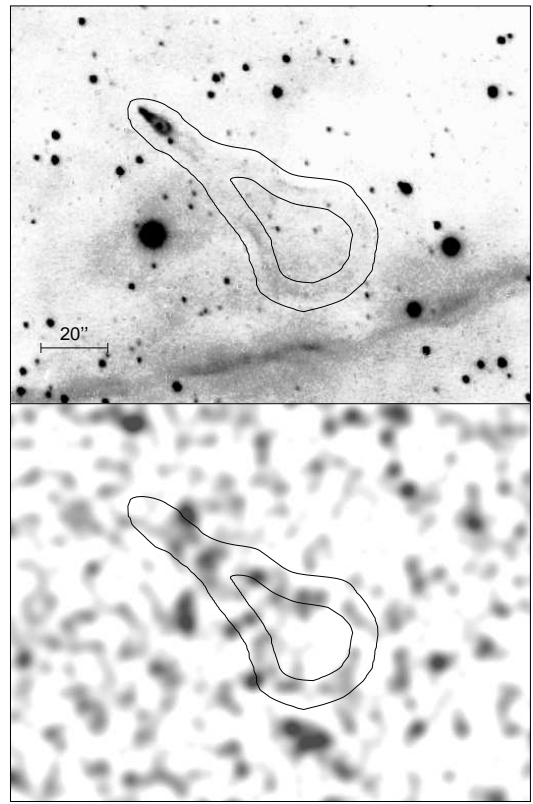}
\caption{Above: H$\alpha$ image of the Guitar Nebula, showing our shaped 
aperture. Below: Aperture over central portion of the smoothed HRI count map.} 
\end{figure}

	Some crude spectral information is available from the PHA energy 
channels in the HRI detector. We find that the excess Guitar counts are soft, 
with a median PHA
channel of 6. The distribution is only slightly harder than that of the mean 
background counts, but also contains an excess of counts in PHA channels 0-1,
suggesting UV flux near the source. This `spectrum' is clearly different 
from some
of the stronger sources identified with bright stars in the optical; these have
median PHA channels of 8-9 and very few counts below channel 5. Unfortunately,
with the very low count rate and lack of calibration, quantitative results 
cannot be extracted from this hardness information.

	The implied X-ray luminosity is sensitive to the parent 
spectrum and the absorption. From the pulsar dispersion measure 
35.3${\rm cm^{-3}/pc}$ and an assumed $n_e/n_H = 0.1$ we infer a column 
density $N_H \approx 10^{21} {\rm cm^{-2}}$. This assumed $n_e/n_H$ is
very similar to that inferred for PSR 2334+61 (Becker, Brazier \& Tr\"umper
1996) at a similar direction and $DM$.
The Balmer decrement in the Guitar nebula itself suggests 
$A_V = 1.65$, corresponding to $N_H = 3 \times 10^{21} {\rm cm^{-2}}$. However, 
because the Balmer ratios in non-radiative shocks are uncertain 
(Chevalier, Kirshner and Raymond 1980; Raymond 1991), we adopt here $10^{21}$ 
for the absorbing column. With this 
$N_H$ and a photon number index of $\alpha=2$ we infer an X-ray flux of 
$1.8 \times 10^{-13} {\rm erg/cm^2/s}$ (0.01-10keV); for 
$N_H = 3 \times 10^{21} {\rm cm^{-2}}$ the flux is 50\% higher. This represents 
$1.7 \times 10^{-2} d_{kpc}^2/I_{45}$ of the spindown luminosity.

\begin{figure}[hist]
\plotone{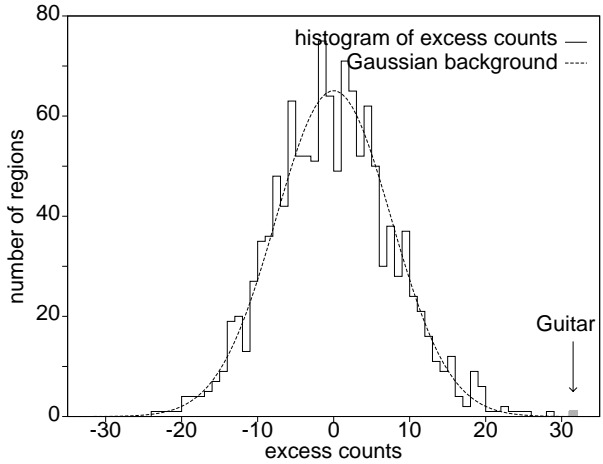}
\caption{Histogram of excess counts in non-overlapping Guitar-shaped 
apertures in the
HRI image. The observed counts at the Guitar position are in the shaded
bin.}
\end{figure}

\section{Geometry of the Guitar Nebula}\label{shape}

	As described in CRL, the guitar nebula is an H$\alpha$-emitting
bow shock trailing the slow spin (P=0.68s) intermediate age ($\tau_c =
1.1 \times 10^6$y) pulsar PSR 2224+65. With an interferometrically determined
proper motion giving a space velocity of $v_p = 8.4 \times 10^7
d_{\rm kpc}/{\rm cos}\,i\,$cm/s (with $i$ the angle between the 
velocity and the plane 
of the sky) and a dispersion measure distance estimate of $d_{\rm kpc} =1.95$
kiloparsecs (Taylor \& Cordes 1993) this object has arguably the highest
space velocity of any known star.

	With this velocity the 
standoff of the pulsar wind bow shock is especially small for this nebula.
The pulsar spindown luminosity ${\dot E} = 1.3 \times 10^{33} 
I_{45} {\rm erg/s}$ is believed to power a relativistic wind which
shocks against the ISM of density $\rho =2.2 \times
10^{-24}n_H\, {\rm g/cm^3}$, giving a momentum balance standoff of the
contact discontinuity at the bow shock nose of
$$
r_0 = ({ {\dot E} \over {4\pi c \rho v_p^2}} )^{1/2} = 
4.7 \times 10^{14} ({ {I_{45}} \over {n_H} })^{1/2}
{\rm cos}\,i/d_{\rm kpc}\,{\rm cm}.
\eqno (1)
$$
The Balmer emission of the pulsar wind nebulae is caused by
neutral ISM atoms drifting into the shocked gas and suffering
collisional excitation or charge exchange into excited states of the post-shock
ions. Accordingly H$\alpha$ arises in a thin sheet behind the
ISM shock, upstream from the contact discontinuity
(Chevalier, Kirshner and Raymond 1980). A detailed model for
the bow shock flow is not available, but Aldcroft, Romani \& Cordes (1992)
estimated that the H$\alpha$ apex lies at $\sim 1.3 r_0$.
This size was unresolved in ground based images of the Guitar.
HST {\it Wide Field Planetary Camera} (WFPC2) H$\alpha$ 
images of the nebula, however, show that the
apex has the classic bow shock shape displayed by the PSR 1957+32
and PSR J0437-4715 nebulae. These data are described elsewhere 
(Cordes {\it et al.} in preparation). Here we note only that while the pulsar,
and hence the standoff angle, are not directly observed in the HST data, a
fit to the bright H$\alpha$ ridge of the bow shock allows us to measure
the size of the bow shock and infer $r_0$. Wilkin (1996) has derived 
convenient analytic expressions for the geometry of a thin momentum-balance 
bow shock. Assuming that the shock standoff at the apex scales proportionally 
for the rest of the bow wave, the scaled version of the momentum balance 
contact discontinuity curve gives the bow shock shape:
$$
\theta_{H_\alpha} (\phi) = 1.3 r(\phi)/d = 1.3
r_0\, {\rm csc}\,\phi \,\sqrt{ 3(1-\phi \,{\rm cot}\,\phi ) }/d
\eqno (2)
$$
where the pulsar is at the origin and position angle $\phi$ is measured from 
the pulsar velocity. The fit determines the origin and the overall scale,
giving an inferred $H\alpha$ standoff angle as $\theta_{H\alpha}(0) =
0.038^{\prime\prime} \pm 0.005^{\prime\prime}$. This corresponds to a 
contact discontinuity
scale $r_0 = 4.3 \times 10^{14} d_{kpc}/{\rm cos}\,i\,$cm which, with (1)
gives an estimate of the pulsar distance as $d_{kpc} = 1.1 {\rm cos}\,i\,
(I_{45}/n_H)^{1/4}$.

	Ground-based Palomar 200$^{\prime\prime}$ images (Figure 1) show 
that the shape of the $H\alpha$ nebula downstream from the {\it HST}
resolved apex can be described as a $\sim 3^{\prime\prime} \times
20^{\prime\prime}$ tube (the `neck' of the Guitar) connecting to
a complex bubble that flares gradually to a diameter of $\sim 30^{\prime\prime}$
some $70^{\prime\prime}$ behind the pulsar. The waist in this bubble
(the Guitar `body') may be due to ISM density variations. The bubble also
does not appear fully closed at the end opposite the pulsar. With
the observed proper motion, the pulsar traverses the Guitar in
$\sim 430$y. CRL give the H$\alpha$ flux of the head of the nebula as
$1.1 \times 10^{-3} {\rm photons/cm^2 /s}$. This line flux implies that 
neutral H atoms enter the nebula head at $1.3 \times 10^{41} d_{kpc}^2 
{\rm atoms/s}$.  With the observed radius of $\sim 1.5^{\prime\prime}$, an ISM 
neutral fraction $X$ and the measured pulsar proper motion, the nebula 
head should
sweep up neutral H at $1.3 \times 10^{41} n_{H} d_{kpc}^3/X {\rm atoms/s}$.
Thus the observed line flux is in good accord with the HST constraints
on the bow shock geometry for $n_{H} d_{kpc}/X \sim 1$.

	The variation in nebula cross section can be understood from 
pressure balance arguments. Behind the pulsar the outflowing 
wind drives the bow shock tail into the ISM; this stalls where the wind
ram pressure balances the ISM static pressure $\sim 10^{-12}P_{-12}\,
{\rm g/cm/s^2}$. Well behind the pulsar the wind impacts at a shallow angle;
a characteristic position angle of $\phi \sim 150^\circ$ gives 
a transverse pressure
$\sim {\dot E}\, {\rm sin^3}(\pi - \phi)/(4\pi \theta_\perp^2 d^2 c)
\sim 2 \times 10^{-12} I_{45} (\theta_\perp^{\prime\prime} d_{kpc})^{-2} 
{\rm g/cm/s^2}$ in a tube of radius $\theta_\perp^{\prime\prime}$ arcseconds.
Thus we expect the ram pressure driven `neck' of the Guitar to stall at 
a width $\theta_\perp \sim 1.4^{\prime\prime} (I_{45}/P_{-12})^{1/2}/d_{kpc}$.
Further downstream it is clear that the shocked pulsar wind is partly
confined in the body of the Guitar. This confinement may be aided by the
increased ram pressure associated with expansion into a dense region
toward the rear of the Guitar, delimited by the horizontal filament in
Figure 1.  Here the isotropic pressure of the shocked 
relativistic plasma $P_{rel} = u/3$ comes into play. If the wind emitted during 
the time the pulsar crosses the Guitar is fully confined in a tube of diameter
$r_\perp$ and there is no cooling, the pressure is 
$ {\dot E}/(3 \pi r_\perp^2 v_p)$. In practice the back of the nebulae
does not appear closed, so perhaps $\sim 0.1$ of this pressure obtains;
even with this pressure the bubble stalls with a characteristic radius 
$\theta_\perp \approx 26^{\prime\prime} (I_{45}/P_{-12})^{1/2}d_{kpc}^{-3/2}$.
We thus have a consistent picture of the Guitar bubble as a classic bow shock
whose ram pressure stalls against the static ISM pressure behind the pulsar;
the backflow in this neck should be mildly relativistic.
Further downstream, the shocked pulsar wind stalls and becomes partly confined.
The shocked wind expands against the ISM inflating the bubble to the
observed $\sim 30^{\prime\prime}$ diameter.

\section{Implications and Bow Shock Flow}\label{Interp}

	We turn now to the interpretation of the lack of X-ray flux from 
PSR 2224+65 and to the X-ray flux from the Guitar nebula body.
The non-detection of the pulsar is not surprising. At $\tau_c=10^6$y
the expected surface temperature of $\sim 3 \times 10^5$K from standard 
cooling models predicts a luminosity of $\sim 6 \times 10^{30} {\rm erg /s}$ 
and no significant HRI flux.  Pulsars of similar age have been detected 
by ROSAT, eg. PSR 1929+10 (Yancopoulos et al. 1994) and PSR 0823+26 
(Sun et al 1993), but the emission from these nearby objects seems to be 
from a small heated polar cap; scaled to PSR 2224+65's distance the
polar cap flux of PSR 1929+10 would contribute $<1$ count in our exposure.

	The two plausible sources of the nebular X-ray emission are the 
shocked pulsar wind and the postshock ISM. In contrast to the bow shocks 
around the slower pulsars PSR 1957+20 and PSR J0437-4715, the large 
pulsar velocity here gives high $T$ in the shocked ISM gas; at the apex of 
the bow wave, the postshock ISM temperature is $3 \times 10^7 d_{kpc}^2$K.
However, the small working surface resulting from the small $r_0$ means 
that the shock luminosity near the Guitar nose is only $\sim \pi r_0^2 
\rho v_P^3/2 \sim 8 \times 10^{29} n_H^{1/2}d_{kpc}^2 {\rm erg/s}$.
Moreover, for a typical ISM abundances and cooling rate $\Lambda$, the 
characteristic cooling time $3kT/(8n_H \Lambda) \sim 8 
\times 10^{13} n_H^{-1}{\rm s}$ is quite long, so little thermal emission 
is expected before adiabatic losses in the post-shock flow become 
significant. The absence of efficient cooling is supported by the 
Balmer-dominated nebular spectrum at the apex (CRL) which indicates that 
the shock is non-radiative (Chevalier, Kirshner and Raymond 1980). Along the 
body of the Guitar, the
continued expansion of the partly confined pulsar wind drives shock waves into
the surrounding medium at ${\bar v} \sim 1.5 \times 10^7 d_{kpc} {\rm cm/s}$. 
The implied shock luminosity along the Guitar body is $L_{sh} \sim 2 
\times 10^{32} d_{kpc}^3 n_H {\rm erg /s }$, with a postshock temperature 
of $\sim 4 \times 10^5 d_{kpc}^2$K.  Although the cooling time for 
$d_{kpc}\sim 1$ is comparable to the flow time through the nebula (so 
that we might expect some thermal emission from the tail of the Guitar), the
effective temperature is too low to produce the observed HRI count rate. 
Even if the distance to PSR 2224+65 is as large as 2kpc, prospects for thermal
emission from the nebula are not much better. With $T_{sh} = 1.6 \times 10^6$K
our observed HRI counts imply a thermal luminosity at 2 kpc of 
$8 \times 10^{32} {\rm erg /s}$ for $N_H = 10^{21}{\rm cm ^{-2}}$ and a 
Raymond-Smith 
thermal spectrum. However at these temperatures the cooling time 
is $\sim 3 \times 10^{12} n_H^{-1}$s, so that radiation in the postshock 
flow is inefficient. Thus unless $d_{kpc}$ and $n_H$ are large,
$N_H$ is small and the cooling rate is significantly enhanced, thermal 
emission cannot account for the observed HRI flux.

	A more promising source is post-shock synchrotron cooling of the 
relativistic pulsar wind.  In the ram-pressure balance region of the bow 
wave, the energy dissipated in this shocked wind is larger than that in 
the shocked ISM by $c/v_P \sim 300$. Observations of the Crab nebula plerion 
suggest that the pulsar wind is dominated by energetic $e^\pm$ with 
$\gamma = 10^7\gamma_7$, $\gamma_7 \sim 0.3$. These electrons
radiate in the postshock field of the pulsar wind, $B_s = 2 
\times 10^{-3} (r_0/r)$G.  Kulkarni \et ~ 1992 were able to use the 
X-ray flux from the millisecond pulsar PSR 1957+20 to limit the fraction
of the pulsar spindown in $e^\pm$ with $\gamma_7 \sim 0.01$ and 
$\gamma_7 \sim 10$,
assuming synchrotron emission at the companion shock and at the bow shock,
respectively. With it's larger field and high velocity, PSR 2224+65's soft 
X-ray flux is most sensitive to intermediate electron energies, similar to 
those in the Crab Nebula.  Including the $r^{-1}$ dependence of the pulsar wind 
$B$ we find that the synchrotron spectrum in the shocked wind has a 
peak energy $E_c \approx 3.5 (r_0/r) \gamma_7^2$keV, so that closer 
than $r_{max} \sim 15 \gamma_7^2 r_0$ the charges will
radiate in the ROSAT band.

	We expect the shocked pulsar wind to have a mildly relativistic 
backflow in the bow shock, with a characteristic flow time at a distance 
$r$ from the pulsar of $\tau_{fl} \sim 3r/c$. This is shorter than the 
synchrotron cooling time under these conditions, so that the efficiency 
is small $\eta \sim \tau_{fl}/\tau_{sy} \sim 3 \times 10^{-3} 
(r_0/r)\gamma_7$. When the peak energy is well above the ROSAT threshold
the observed flux will decrease somewhat; since the synchrotron emissivity
at energy $E$ 
scales as $\sim E^{1/3} {\rm exp}(-E/E_c)$, the fraction of the radiation 
in the ROSAT band has the weak dependence $(0.3{\rm keV}/E_C)^{1/3}\sim 
0.4 (r/r_0)^{1/3} \gamma_7^{-2/3}$. Thus if a fraction $f_\gamma$ of 
the spindown energy is carried off by charges of Lorentz factor $\gamma$, 
then including the cooling efficiency and the fraction of the spectrum 
in the ROSAT band, we get an observed X-ray luminosity from the wind 
shock within position angle $\phi$ of
$$
L_x \sim 10^{-3} f_\gamma {\dot E} (1- {\rm cos}\phi ) 
 \left [ { {r_0} \over {r(\phi)} } \right ]^{2/3}\gamma_7^{1/3}
$$
where $r(\phi)$ is determined from (2). Thus the emission should be 
concentrated toward the bow shock apex. Because the wind cannot cool 
effectively before the relativistic backflow carries it towards the body 
of the Guitar, the total luminosity is small.  With an $\alpha=2$ spectrum 
and $N_H=10^{21}{\rm cm^{-2}}$ at 1kpc, we expect 2 counts from the
shocked pulsar wind at the apex of the flow.

	Above we have assumed that the field in the pulsar wind itself 
dominates $B$ in the post-shock region and that the flow is not confined. 
However, it appears that the pulsar wind is partly confined in the nebula. 
In addition turbulence in the shocked pulsar wind may increase $B$ up 
to equipartition values. At the shock apex $B_{eq} \ll B_s$, so that our 
previous estimates hold. However, $\sim 30r_0$
($\sim 1^{\prime\prime}$) behind the pulsar the ram pressure associated 
with the sideways expansion of the guitar body at 
$1.5 \times 10^7 d_{kpc} {\rm cm/s}$
begins to exceed $B^2/4\pi$ in the shocked pulsar wind. Thus as the neck of the
Guitar merges into the body we expect that an equipartition field can
reach $B_{eq}  \sim 8 \times 10^{-5} d_{kpc}n_H^{1/2}$G. The characteristic 
energy in this field $E_c \sim 0.15 \gamma_7^2$keV is large enough to 
contribute significant ROSAT emission. More importantly, if the shocked 
wind is partly confined in this region, we expect $\tau_{fl} \sim r/v_p 
\sim 5 \times 10^9$s. For $d_{kpc}=1$ we get a synchrotron cooling time 
$\tau_{sy} \sim 9 \times 10^9 \gamma_7 ^{-1}$s
and about 30\% of the shock energy radiated in the ROSAT band for $\gamma_7=1$.
This gives an observed X-ray flux $0.3 {\dot E} f_\gamma \tau_{fl}/\tau_{sy}$.
Thus to produce the observed X-rays, we require $f_\gamma \sim 0.1$ in 
$\gamma_7 > 1$ particles. With the increased $B_{eq}$ inferred at 2kpc, cooling
is even more efficient and we again infer $f_\gamma \approx 0.1$.

\section{Conclusions}

   The X-ray emission observed from the bow shock created by PSR 2224+65 
represents $\sim 0.02 d_{kpc}^2/I_{45}$ of it's spindown power. Although 
the pulsar's large velocity ensures that gas is heated to X-ray emitting 
temperatures near the bow wave, the cooling in the post-shock ISM flow is 
inefficient. Little of the observed flux can be contributed by the shocked 
ISM. The location of the X-ray emission significantly downstream from the 
pulsar suggests that the most likely source of the flux is the shocked 
pulsar wind. However, to account for the observed flux we require this 
wind to be partly confined in the Guitar body and that turbulence
in the shocked wind flow can amplify the magnetic field to near equipartition 
vales.  If these conditions obtain, the observed flux implies that 
$> 0.1$ of the pulsar spindown power is being emitted in $e^\pm$ with 
$\gamma \sim 10^7$. Our data then support the idea that a pair plasma with
$\gamma \sim 10^6 - 10^7$ is a dominant component of the pulsar wind for
normal non-recycled pulsars. 
Such a wind is in reasonable accord with models of 
pulsar magnetosphere acceleration, but the pairs may also be accelerated 
in the post-shock region itself. In contrast the conditions inferred for
the soft X-ray emission from PSR 1929+10's shock by Wang, \et ~suggest that
$\gamma > 10^8$ pairs are responsible for the PSPC photons; it is perhaps
significant that the flux from these pairs represents only 
$10^{-4}$ of PSR 1929+10's spindown luminosity.

	A more complete description of thermal and non-thermal emission in 
the Guitar nebula will require detailed hydrodynamic models of the backflow 
in the postshock region. For example, Raga, Cabrit \& Canuto (1995) find that 
there is mixing of the shocked fluids in a bow shock flow. While the 
presence of the relativistic pulsar wind means that their sums are not 
directly applicable to PSR 2224+65, such mixing could affect the estimates 
of cooling time in the postshock flow. Such modeling coupled with future
high throughput X-ray observations could give important constraints on the
pulsar wind in this unique system.

\begin{acknowledgements} 
	This work was supported in part by NASA grants NAG
5-2706 and NAGW-4526 (RWR), and NSF grant AST 92-18075 (JMC).
\end{acknowledgements}

\end{document}